\documentclass[prc,aps,nofootinbib,showkeys,showpacs,twocolumn]{revtex4}   
\usepackage{epsfig}  
\usepackage{graphicx}  
\usepackage{color}  
\begin{document}  
  
\title{Pairing dynamics in particle transport}

\author{Guillaume Scamps}  
 \email{scamps@ganil.fr}  
\affiliation{GANIL, CEA/DSM and CNRS/IN2P3, Bo\^ite Postale 55027, 14076 Caen Cedex, France}  
\author{Denis Lacroix} \email{lacroix@ganil.fr}  
\affiliation{GANIL, CEA/DSM and CNRS/IN2P3, Bo\^ite Postale 55027, 14076 Caen Cedex, France}  
\author{G.F.~Bertsch}  \email{bertsch@uw.edu}  
\affiliation{Institute for Nuclear Theory and Department of Physics, University of Washington, Seattle, Washington 98195, USA}  
\author{Kouhei Washiyama} \email{kouhei.washiyama@ulb.ac.be}  
\affiliation{Physique Nucl\'eaire Th\'eorique, CP229, Universit\'e Libre de Bruxelles, B-1050 Bruxelles, Belgium}  
\affiliation{GANIL, CEA/DSM and CNRS/IN2P3, Bo\^ite Postale 55027, 14076 Caen Cedex, France}

\begin{abstract}  
We analyze the effect of pairing on particle transport in time-dependent  
theories based on the Hartree-Fock-Bogoliubov (HFB) or BCS approximations. 
The equations of  
motion for the HFB density matrices are unique and the theory respects  
the usual conservation laws defined by commutators of the conserved   
quantity with the Hamiltonian.  In contrast, the theories based on the 
BCS approximation are more problematic.  
In the usual  
formulation of TDHF+BCS, the equation of continuity is violated and  
one sees unphysical oscillations in particle densities.  This can  
be ameliorated by freezing the occupation numbers during the   
evolution in TDHF+BCS, but there are other problems with the BCS that
make it doubtful for reaction dynamics.  We also compare different numerical implementations 
of the time-dependent HFB equations.   
The  equations of motion for the $U$ and $V$ Bogoliubov transformations are not  
unique, but it appears that the usual formulation is also the most  
efficient.  Finally, we compare the time-dependent HFB 
solutions with  numerically exact solutions of the two-particle 
Schr\"odinger equation.
Depending on the treatment of the initial state, the HFB dynamics
produces a particle emission rate at short times similar to that of
the Schr\"odinger equation.  At long times, the total particle emission
can be quite different, due to inherent mean-field approximation of
the HFB theory.
\end{abstract}  
\maketitle  
  
\section{Introduction}  
  
Pairing is essential to the global description of nuclear ground-state and  
low excited state properties; the Hartree-Fock-Bogoliubov (HFB) and  
Hartree-Fock augmented by BCS (HF+BCS) theories are in common use to treat  
the pairing degrees of freedom \cite{Ben03}. Also in nuclear reactions,  
many phenomena are expected to be influenced by pairing correlations:  
collective motion, fusion, fission, transfer reactions and nuclear break-up.   
The obvious candidate theory to treat these effects is the Time-Dependent  
Hartree-Fock Bogoliubov (TDHFB) theory \cite{Sim10}, and there has been much effort in  
the last decade to apply it.  However, the TDHFB theory turns out to be much  
more complicated to implement than the corresponding time-dependent  
Hartree-Fock theory, and the applications have been mainly performed in its small  
amplitude limit, the Quasi-particle RPA (QRPA)  
\cite{Hag01,Mat01,Gra01,Kha02,Paa04,Ter06,Nak04,Per11}.  However, most  
of the phenomena quoted above are far from small amplitude excursions from the  
ground state, and transport theories able to treat Large Amplitude  
Collective Motion (LACM) are mandatory.   
Recently, several groups have  
applied the TDHFB \cite{Has07, Ave08, Ste11} to nuclear dynamics.  An   
approximate version of theory,  
called TDHF+BCS,  has also been considered \cite{Eba10}. We will  
discuss its properties as well.    
Most recent applications have been to small-amplitude  
collective motion in nuclei, where the theory is equivalent to the  
quasiparticle random-phase approximation (QRPA).   Still, it is important to understand and   
solve the theory here as a  
first step towards treating large-amplitude motion.  
     
The aim of the present article is first to present from a rather general  
point of view different dynamical theories that incorporate pairing  
correlations.  We find that the TDHFB has many good properties, but it  
is difficult to find further simplifying approximations.  We find that the   
TDHF+BCS approximation   
leads to a break-down of the continuity equation. This failure might induce  
serious difficulties in the description of physical processes.  The   
second aim of the paper is to test various implementations of   
the pair theory in a model problem.  For this purpose, we examine   
a one-dimensional model of particle evaporation for which we also  
have a numerical exact solution.  
   
\section{Formalism}  
  
The TDHFB theory and the TDHF+BCS approximations have been recently applied to  
nuclear physics in Ref. \cite{Has07, Ave08, Ste11, Eba10}.  In this section  
we will briefly summarize the main features.  
  
\subsection{The TDHFB theory}  
\label{sec:tdhfb}  
  
The main equations of TDHFB theory, Eq. (\ref{eq:evol_rho}-\ref{eq:evol_kappa}) below, can be derived in at   
least two different ways. One way is from  
the general variational principle,   
\begin{eqnarray}  
{S} & = & \int_{t_i}^{t_f} \langle \Psi(t) | i\hbar \partial_t - H | \Psi(t)   
\rangle dt,  
\label{eq:varia}   
\end{eqnarray}  
see e.g.  Ref. \cite{Bla86}.  
Here $H$ denotes the Hamiltonian and $| \Psi \rangle$ is a HFB wave  
function.   The equations may also be derived by demanding that the operators for the  
ordinary and anomalous densities, $\hat \rho_{ij} = a^\dagger_j a_i$  
and $\hat \kappa_{ij} = a_j a_i$ respectively, satisfy Ehrenfest's  
theorem:  
\begin{eqnarray}  
i \hbar \partial_t \langle\Psi| a^\dagger_j a_i|\Psi\rangle   
= \langle \Psi|[a^\dagger_j a_i ,H ]|\Psi\rangle, \label{OBDM}\\  
 i \hbar \partial_t \langle\Psi|  a_j a_i | \Psi\rangle   
= \langle\Psi| [  a_j a_i ,{H} ]|\Psi\rangle . \label{OBPT}  
\end{eqnarray}  
In the following, we will further assume that $H$ is a two-body hamiltonian,  
\begin{eqnarray}  
H = \sum_{i j } h^0_{ij} a^\dagger_i a_j + \frac{1}{4} \sum_{ijkl} \bar{v}_{ijkl} a^\dagger_i a^\dagger_j a_l a_k.  
\end{eqnarray}  
where $\bar{v}$ denotes the anti-symmetric two-body matrix elements.  The  
derived TDHFB equations are:  
\begin{eqnarray}  
i \hbar \frac{d}{dt} \rho & = &  h  \rho - \rho h  + \kappa \Delta^*  - \Delta  \kappa^*,
\label{eq:evol_rho}\\
i \hbar \frac{d}{dt} \kappa & = & h \kappa + \kappa h^* + \Delta (1-\rho^*) - \rho \Delta.  
\label{eq:evol_kappa}  
\end{eqnarray}   
Here $\rho$, $\kappa$,  $h$ and $\Delta$ are all matrices of dimension equal  
to that of the single-particle space.  
The matrices $h$ and $\Delta$ are the mean-field and pairing field of  
the Hamiltonian, defined as  
\begin{eqnarray}  
\label{eq:h}  
h_{ij} &=& h^0_{ij} + \sum_{kl} \bar{v}_{iljk} \rho_{kl}, ~~~\Delta_{ij} = \frac12 \sum_{kl} \bar{v}_{ijkl} \kappa_{kl}.  
\end{eqnarray}

The dynamical equation can be recast in a more compact form by introducing   
the generalized density matrix $\cal R$ and generalized single-particle  
hamiltonian $\cal H$:   
\begin{eqnarray}  
\cal R =   
\left(  
\begin{array} {cc}  
 \rho &  \kappa \\  
 -\kappa^*  &  1-\rho^* \\  
\end{array}   
\right),  
\qquad  
\cal H =   
\left(  
\begin{array} {cc}  
 h &  \Delta \\  
 -\Delta^*  &  -h^* \\  
\end{array}   
\right).  
\end{eqnarray}  
With these definitions the equation of motion becomes~\cite[Eq. 9.61a]{Bla86}  
\begin{eqnarray}  
i \hbar \frac{d}{dt} {\cal R} = \left[ {\cal H,R} \right].  
\label{eq:dRdt}  
\end{eqnarray}  
This generalizes the usual TDHF picture by replacing the one-body   
density by ${\cal R}$. Similarly to the TDHF case, the generalized density   
satisfies ${\cal R}^2 = {\cal R}$  
and has only eigenvalues equal to zero and one~\cite{RS,Bla86}.   
  
Going back to ordinary Hartree-Fock theory, it is computational  
advantageous  
to factorize the density matrix and express it as a sum over the contributions  
from occupied orbitals to obtain equations of motion for the individual  
orbitals. There is no obvious  
advantage for the factorization in TDHFB because all of the single-particle  
orbitals in Fock space contribute to the generalized density matrix $\cal R$.   
Nevertheless, the factorization is usually applied to obtain the actual  
equations to be solved numerically.  To write equations in this form, one needs an  
explicit form of the Bogoliubov transformation,  
\begin{eqnarray}  
\beta_\alpha   = & \sum_{i} U^*_{i \alpha} a_i + V^*_{i\alpha} a^\dagger_i.  
\label{eq:bogogen}  
\end{eqnarray}  
The density matrices are expressed as $\rho = V^*V^T$ and $\kappa = V^*U^T$,  
and the generalized density matrix is  
\begin{eqnarray}  
{\cal R} &=  
	\left(  
	\begin{array} {c}  
		 V^*\\  
		 U^* \\  
	\end{array}   
	\right)  
	\left(  
	\begin{array} {cc}  
		 V^T  &  
		 U^T \\  
	\end{array}   
	\right)  
	\label{eq:Rmatrix}.  
\end{eqnarray}  
One can then easily see that Eq. (\ref{eq:dRdt}) will be satisfied if we require  
the $\{U , V \} $ matrix be a solution of  
\begin{eqnarray}  
i\hbar \frac{d}{dt}\left(  
	\begin{array} {c}  
		 U \\  
		 V \\  
	\end{array}   
	\right) &=& 
	\left(  
	\begin{array} {cc}  
		 h &  \Delta \\  
		 -\Delta^*  &  -h^* \\  
	\end{array}   
	\right)  
	\left(  
	\begin{array} {c}  
		 U \\  
		 V \\  
	\end{array}   
	\right)  
	\label{eq:wave1}.  
\end{eqnarray}  
The numerical solution of the TDHFB equations are usually carried out  
in this representation \cite{Ave08,Ste11}.  However, it should be remembered  
that there are redundant variables in the $\{U , V \}$ representation  
corresponding to unitary transformations of the quasiparticle basis,  
and in fact Eqs. (\ref{eq:wave1}) are not unique.    
  
We may derive another form of the TDHFB equations   
as follows.  The wave function $|\Psi\rangle$ at any time $t$ is   
the quasi-particle vacuum associated with the Bogoliubov transformation  
that transforms the physical vacuum to $|\Psi(t)\rangle$.  In that  
representation, the Hamiltonian has zero-, two-, and four-quasiparticle  
terms that can act on $|\Psi(t)\rangle$~\cite{RS}.  Neglecting   
the four-quasiparticle excitation amplitudes, the result is  
\begin{eqnarray}  
H | \Psi(t) \rangle &\simeq& H'(t)  | \Psi (t)\rangle  \nonumber \\  
&=& \left[ \langle H \rangle + \frac{1}{2}  
\sum_{\alpha \beta } H^{20}_{\alpha \beta}  \beta^\dagger_\alpha(t) \beta^\dagger_\beta(t) \right] | \Psi (t)\rangle   
\label{eq:hamilt_bogo}  
\end{eqnarray}  
with \cite[Eq. (E.22)]{RS}  
\begin{eqnarray}  
H^{20} &=& U^\dagger h V^* - V^\dagger h^T U^* + U^\dagger \Delta U^* - V^\dagger \Delta^* V^*.  
\end{eqnarray}   
  
According to the Thouless theorem, any state of the form $(1 + \sum_{\alpha\beta}  
Z_{\alpha\beta}\beta^\dagger_\alpha\beta^\dagger_\beta)|\Psi\rangle$ can be expressed as a new quasiparticle  
vacuum~\cite{RS}.  To lowest order in $Z$, the Bogoliubov transformation to the new vacuum from the   
physical vacuum is given by   
\begin{eqnarray}  
\left(  
	\begin{array} {cc}  
		 U' &  
		 {V^*}' \\  
	\end{array}   
	\right) &=&   
	\left(  
	\begin{array} {cc}  
		 U  &  
		 V^* \\  
	\end{array}   
	\right)  
	\left(  
	\begin{array} {cc}  
		 1 &  Z^* \\  
		 Z^*  &  1 \\  
	\end{array}   
	\right)  
	\label{eq:wave3} \end{eqnarray} We can thus derive an equation of  
motion by demanding that the changes in $U,V$ just match the two quasiparticle  
excitations generated by $H$.  After a lengthy but straightforward  
derivation, it may be shown that the corresponding equations of motion  
for $U$ and $V$ can be written as:   
\begin{eqnarray} \left\{  
\begin{array} {lll}  
 i \hbar \partial_t U & = &  \rho h^\dagger U - \kappa h^* V \\  
 &-& \kappa \Delta^* U + \rho \Delta V  
\\  
\\  
 i \hbar \partial_t V  & = & -(1-\rho^*)h^*V - \kappa^* h^\dagger U  \\  
 &-& \kappa^* \Delta V - (1-\rho^*)\Delta^* U    
 \end{array}  
\right.  
.  
\label{eq:wave2}  
\end{eqnarray}   
These equations differ from (\ref{eq:wave1}) but nevertheless   
lead to the same TDHFB equation for the generalized density.

\subsection{The TDHF+BCS approximation}  
\label{sec:tdhfbcs}  
  
The TDHF+BCS treatment of pairing dynamics is motivated by the simple  
form the wave function has in the BCS   
approximation,  
\begin{eqnarray}  
|\Psi\rangle =\prod_{k>0} \left( u_k + v_k a^\dagger_k a^\dagger_{\bar  
k}\right)|\,\rangle.  
\end{eqnarray}  
The TDHF+BCS approximation may be derived from a   
variational principle \cite{Blo76} or by an approximate  
reduction of the TDHFB equations \cite{Eba10}.  For the reduction  
of the TDHFB equations, we first note that wave function can be put into BCS   
form at any fixed time by transforming the $U,V$ matrices to the canonical  
basis.  In that basis,  $\rho$ is diagonal and $\kappa$ matrix is zero except for one  
element on each row (or column) representing the pair $i \bar i$.  Assuming  
that the $\Delta$ matrix has the same structure as $\kappa$,  Ref.  
\cite{Eba10}  
shows that the TDHFB time evolution preserves the same canonical structure  
with orbitals that evolve by the mean field Hamiltonian,  
\begin{eqnarray}  
\label{eq:phi-mf}  
i\hbar \partial_t | \varphi_k \rangle = h | \varphi_k \rangle.  
\end{eqnarray}  
where $h$ has been defined in Eq. (\ref{eq:h}).  
The equations of motion for $\rho$ and $\kappa$ in this time-dependent  
basis are\footnote{Note that these equations slightly differ from those from \cite{Eba10}, due to the definition of $\kappa_k$ here. }  
\begin{eqnarray}  
i \hbar \frac{d}{dt} n_k &=&   
\Delta_{k} \kappa_{k}^* - \Delta_{k}^* \kappa_{k}, \\  
i \hbar \frac{d}{dt} \kappa_k &=&   
 + \Delta_{k} (1-2n_k).\label{eq:k-bcs}  
\end{eqnarray}  
Here $n_k$, $ \kappa_k $ and $\Delta_{k}$ are short-hand notations for   
$\rho_{kk}$, $\kappa_{ k  
\bar k}$ and $\Delta_{k\bar k}$ respectively.  
  
One technical point should be mentioned.  When Eq. (\ref{eq:phi-mf}) is  
integrated, there is an irrelevant phase factor   
$\exp( -i\int^t\langle \varphi_k(t') |h_{HF}(t') |  
\varphi_k(t') \rangle d t')$  
introduced into the time-dependent orbitals.  For computational   
reasons the phase is removed by integrating   
\begin{eqnarray}  
i\hbar \partial_t | \varphi_k \rangle = (h[\rho]-\eta_k) | \varphi_k \rangle  
\end{eqnarray}  
instead of  Eq. (\ref{eq:phi-mf}), with   
$\eta_k(t) = \langle \varphi_k(t) |h_{HF} |  
\varphi_k(t) \rangle $.  At the same time, Eq. (\ref{eq:k-bcs}) is replaced  
by   
\begin{eqnarray}  
i \hbar \frac{d}{dt} \kappa_k &=&   
\kappa_k ( \eta_k + \eta_{\overline{k}} ) + \Delta_{k} (1-2n_k).  
\end{eqnarray}  
Finally, we mention that   
the TDHF+BCS approximation was found to work well with a separable   
pairing interaction and in the small amplitude  
limit  \cite{Eba10}.  
  
\subsection{Conservation laws and equation of continuity}  
  
Since the TDHFB density matrix satisfies Ehrenfest's theorem, it is trivial  
to show that the conservation laws for one-body observables are respected by the   
TDHFB dynamics.  It was also shown that conservation laws for important observables  
such as particle number are satisfied in TDHF-BCS \cite{Eba10}.   
However, for transport we are interested in local conservation laws as  
well.  In particular, if the interaction is local the coordinate-space  
density $n(x,t)$ should satisfy the equation of   
continuity,   
\begin{eqnarray}  
\frac{d n (x,t)}{dt} & = & - {\vec \nabla} \cdot \vec j(x,t)  
\label{eq:continuity}  
\end{eqnarray}   
where $\vec j(x)$ is the particle current.  Assuming a local interaction,  
Eq. (\ref{eq:continuity})  
may be derived from Ehrenfest's theorem,  
evaluating the commutator on the right hand side as  
\begin{eqnarray}  
 {\vec \nabla} \cdot \vec j(x) = \langle [\hat n(x),H]\rangle = -  
 \frac{\hbar^2}{2m} \langle [ \hat n (x),\nabla^2   ] \rangle.  
\label{eq:continuity2}  
\end{eqnarray}   
This is sufficient to guarantee that TDHFB obeys the equation of continuity  
under the stated condition.  Unfortunately, this is not true for the  
TDHF+BCS dynamics.    
  
Within TDHF+BCS, the local density is given by  
\begin{eqnarray}  
n(x,t) & = & \sum_i n_i(t) |\varphi_i(x,t)|^2,  
\end{eqnarray}   
and its evolution satisfies  
\begin{eqnarray}  
\frac{d n(x,t)}{dt}  & = &   
\sum_i n_i (\varphi_i^*(x,t)  \partial_t \varphi_i (x,t ) )+\varphi_i(x,t)  \partial_t \varphi^*_i (x,t ) ) \nonumber \\  
&+&\sum_i  |\varphi_i(x,t)|^2 \partial_t n_i(t) 
\end{eqnarray}   
The first terms on the left are just the evolution of the orbitals under  
a mean-field potential, and so the same reduction applies as in Eq.   
(\ref{eq:continuity}).  The result is  
\begin{eqnarray}  
\frac{d n(x,t)}{dt}  & = & - {\vec \nabla}\cdot \vec j (x,t)  
+ \sum_i  |\varphi_i(x,t)|^2 \left( \frac{dn_i (t)}{dt} \right).  
\label{eq:contbcs}  
\end{eqnarray}     
Thus continuity cannot be guaranteed unless the occupation numbers  
are fixed.  We will see below that TDHF+BCS can produce unphysical   
density oscillations when the occupations are allowed to vary.

\section{Application to particle evaporation}  
  
\begin{figure}[htbp]  
 \begin{center}  
 \hspace*{.1cm}\includegraphics[width=0.5\linewidth]{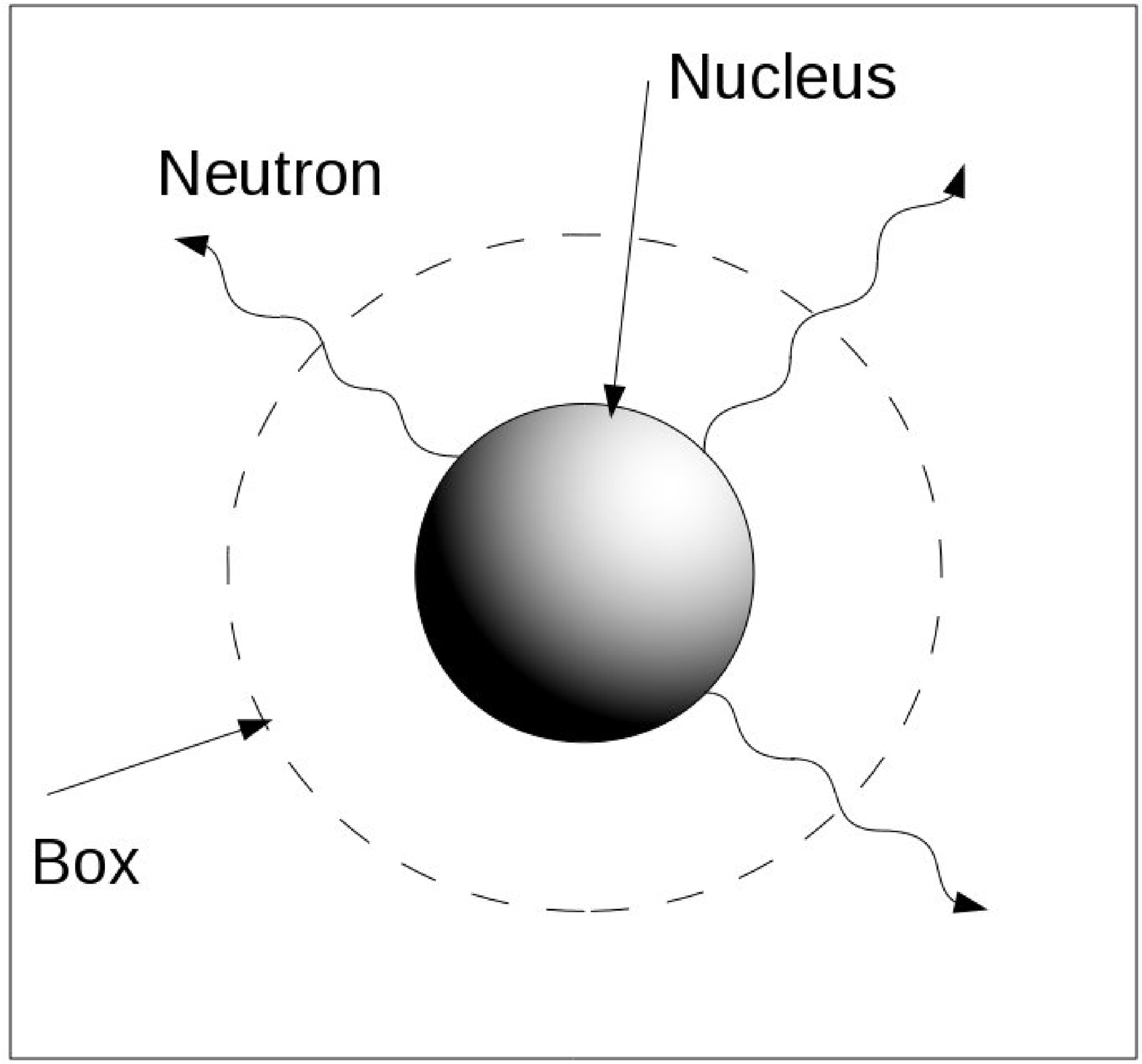} \\  
 \includegraphics[width=0.7\linewidth]{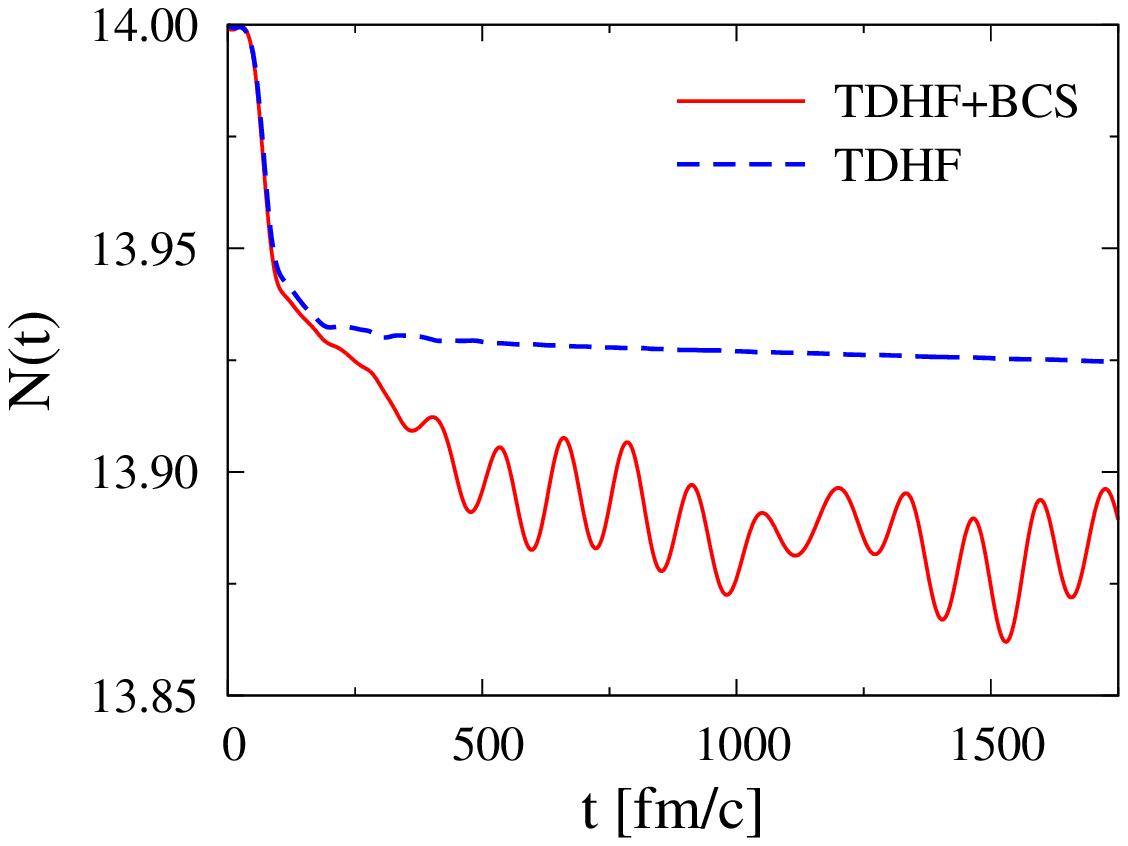}  
 \end{center}  
  \caption{Top: Schematic illustration of neutron evaporation from   
  \textcolor{black}{a nucleus of O$^{22}$ excited by a monopole boost at t=0}.  
 Bottom: Number of neutron inside a  
  sphere of size $10$ fm around the nucleus as a function of time obtained   
  with TDHF and TDHF+BCS (from \cite{Was11}). }  
  \label{fig:intro1}  
 \end{figure}

 Recently two of us (DL and KW) began investigating the effect of pairing on particle  
 evaporation, and obtained the results   
shown in Fig. \ref{fig:intro1}.  Skipping over the details, the number of   
particles escaping an initially excited  
 nucleus is shown as a function of time using either the 3D-TDHF code of  
 Ref.  \cite{Kim97,Sim01,Was08} or an upgraded version including pairing  
 using the TDHF+BCS theory proposed in Refs.  \cite{Blo76,Eba10}.  As we can  
 see, the standard mean-field calculation presents the expected long time  
 decay due to particle evaporation\cite{Van88}.  When pairing is  
 included, the number of particles in the nucleus first decays and then  
 starts to oscillate.  Clearly, this result is unphysical.  It was this  
unphysical result that motivated us to undertake the present more general  
study.  For the present article, we consider a more simplified Hamiltonian that  
permits us to compare a number of approximations with each other and with  
a numerically exact solution. In the present article, we investigate whether the observed problem is systematic in theories   
 where pairing is included or if it comes from the specific treatment of pairing in the TDHF+BCS approximation using   
 zero range interaction. Our study is also the occasion to benchmark different theories, TDHF+BCS and TDHFB,   
 to describe particle emission.   

\subsection{A one-dimensional model}

For comparing the different treatments of pairing dynamics,   
we consider a one-dimensional system composed of $N$ particles in a box  
with $x$ in the range $-X_{\rm max} < x < X_{\rm max} $   
and a Hamiltonian of the form  
\begin{eqnarray}  
H & = & \sum_i^N \left\{ \frac{p_i^2}{2m} + U(x_i) \right\} \nonumber \\  
&+& \sum_{i<j}^{N(N-1)/2} v(x_i - x_j) [1 - P_{\sigma_i \sigma_j}].  
\end{eqnarray}  
Here, $P_{\sigma_i \sigma_j}$ denotes the spin-exchange operator. The  
potential $U(x)$ is taken to be a Woods-Saxon well centered at the  
origin:  
\begin{eqnarray}  
U(x) & = & \frac{U_0} {1+ \exp [(|x|-X_0)/a]}.  
\end{eqnarray}   
The two-body interaction $v(x - x')$ is taken to be a finite-range  
Gaussian  
\begin{eqnarray}  
v(x - x') & = & v_0 \exp\left(- \frac{(x-x')^2}{2\sigma^2_0} \right)  
 \end{eqnarray}   
In the limit where the range $\sigma_0$ goes zero,   
$v(x - x')$ is a contact interaction and   
our model is similar to the model considered in Ref. \cite{Ton06}   
to analyse the onset of vortices in rotating Fermi gas using TDHFB. The  
advantage of a finite range is that it does not have to be renormalized  
for use in BCS or HFB.  
   
The TDHFB is formulated in a Fock space and the space has finite dimension  
in numerical implementations.  
Our particle creation and annihilation operators  
$\psi_\sigma^\dagger,\psi_\sigma$   
are defined on a uniform mesh of points $\{x\}$ with spacing $\Delta x$; $\sigma =  
\uparrow$ or $\downarrow$ is the spin label.    
Then we can write the quasiparticle transformation as  
\cite{Ton06} 
\begin{eqnarray}  
\beta^\dagger_{\alpha }  &=& \Delta x\sum_{x}   
\left(   
u_\alpha ( x , t )  \psi^\dagger_{\uparrow} (x) +  v_\alpha (x,t)  
  \psi_{\downarrow} (x)\right). \\  
\beta^\dagger_{\alpha' }  &=& \Delta x\sum_{x}   
\left(   
u_{\alpha'} ( x , t )  \psi^\dagger_{\downarrow} (x) +  v_{\alpha'} (x,t)  
  \psi_{\uparrow} (x)\right).  
\end{eqnarray}  
In the following, we will use the convention $\Delta x\sum_{x} \rightarrow   
\sum_{x} $ and not distinguish between the quasiparticle sets $\alpha$ and 
$\alpha'$,
 with the property $u_{\alpha'} ( x , t )  = u_{\alpha} ( x , t ) $ and $v_{\alpha'} (x,t)  = -v_{\alpha} (x,t) $.  
The discretized time-dependent equations in version Eq. 
(\ref{eq:wave1}) of TDHFB take   
the explicit form  
\begin{eqnarray}  
i\hbar \frac{\partial}{\partial t} u_\alpha(x,t)& = & \left\{  
-\frac{\hbar^2 \Delta^{(2)}_x}{2m(\Delta x)^2} + U(x) + \Gamma(x) \right\}  u_\alpha(x,t) \nonumber \\  
&-& \sum_{x'}   
\Delta(x,x')v_\alpha(x',t) \label{eq:ualphat}  
\end{eqnarray}   
and   
\begin{eqnarray}  
i\hbar \frac{\partial}{\partial t} v_\alpha(x,t)& = & -\left\{   
-\frac{\hbar^2 \Delta^{(2)}_x}{2m(\Delta x)^2} + U(x) + \Gamma^*(x) \right\}  v_\alpha(x,t) \nonumber \\  
&& - \sum_{x'}   
\Delta^*(x,x') u_\alpha(x',t) .  
\end{eqnarray}   
with   
\begin{eqnarray}  
\Gamma(x) &=& \sum_{x'} v(x -x') \rho(x',x'), \\   
\Delta(x,x') &=&  v(x-x') \kappa(x,x').  \label{eq:valphat}  
\end{eqnarray}  
Here $\Delta^{(2)}_x$ is the second-difference operator,   
$\Delta^{(2)} \phi(i) = \phi(i+1)-2 \phi(i) + \phi(i-1)$. 
 
The normal and anomalous density matrix are given by   
\begin{eqnarray}  
\rho(x,x) & = &  \sum_\alpha |v_\alpha ( x , t )|^2 \\  
\kappa(x,x') & = & \sum_{\alpha} v^*_{\alpha}(x)u_{\alpha}(x').   
\end{eqnarray}

\subsection{Exact solution for the two particle case}  
  
One interesting aspect of the model considered here is that for two particles it can be solved exactly numerically.   
Indeed, assuming that the system is a spin singlet, the two-body wave-function reads:  
\begin{eqnarray}  
\Phi(x_1,\sigma_1,x_2,\sigma_2) = \frac{1}{\sqrt{2}} \left( \delta_{\sigma_1 \uparrow} \delta_{\sigma_2 \downarrow} - \delta_{\sigma_1 \downarrow} \delta_{\sigma_2 \uparrow} \right) \phi(x_1,x_2),  
\label{eq:phi2}  
\end{eqnarray}  
where $\phi(x_1,x_2)$ is a symmetric function that satisfies the Schr\"odinger equation:  
\begin{eqnarray}  
i \hbar \frac{d}{dt}\phi(x_1,x_2)= \left[ h^0_1 + h^0_2 + v(x_1-x_2) \right] \phi(x_1,x_2),  
\end{eqnarray}   

Since the discussion here might be applied not only to nuclear systems but
also to other field of physics like condensed matter or atomic physics, we
consider here reduced units.  The length, time-scale and energy scale given
below are respectively written in units of $\Delta x$, $m \Delta x^2/\hbar$
and $\hbar^2/(m \Delta x^2)$ where $\Delta x$ is the discretization mesh
step. Accordingly, all quantities below will be presented without specific
units.  The parameters of the central potential are set to
$a=2$, $X_0=4.5$ and $\sigma_0 =
2.5$ and the initial harmonic constraint is taken as  $\lambda = 6.173 \times
10^{-4}$. Three interaction strength $v_0$ equal to $-1.096 \times 10^{-2}
$, $-3.344 \times 10^{-2}$, and $-6.280 \times 10^{-2}$ are considered. The
three cases will be referred respectively to case (a), (b) and (c) below. 
In each case, the depth of the Woods-Saxon potential has been adjusted to
get the same binding energy $E=-2.2 \times 10^{-2}$, leading to $U_0=-2.7
\times 10^{-2}$, $-1.929 \times 10^{-2}$ and $-7,716\times 10^{-3}$
respectively. For cases (a) and (b) the interaction is below the
strength needed for a condensate in the HFB or BCS theory at a mean particle
number of two.
  
\begin{figure}[htbp]  
  \includegraphics[width=1.0\linewidth]{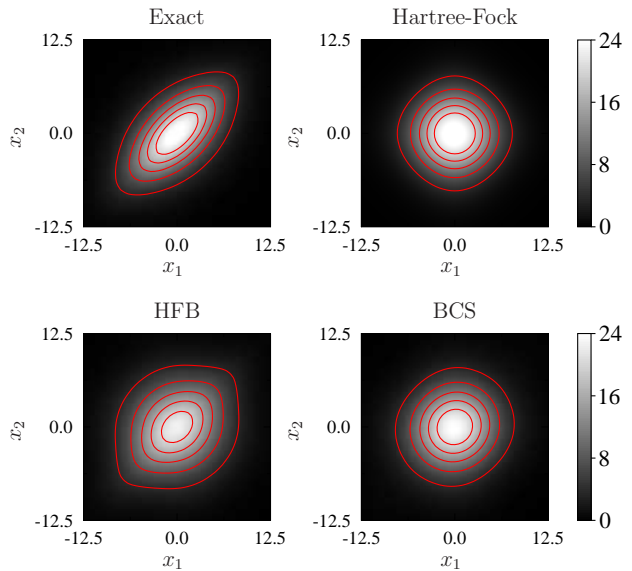}  
  \caption{\textcolor{black}{ $S=0$ component of the two body density matrix   
  $\rho^{(2)}(x_1 \uparrow , x_2 \downarrow)$ in $10^{-3}$ (unit of length)$^{-2}$  at time $t=0$ for the four theories studied here.   
This figure correspond to the set of parameters (c) (see text).}   
 }  
  \label{fig:correlation}  
 \end{figure}

An illustration of the two-body density matrix obtained in different level of approximation for the case (a) 
are shown in Fig. \ref{fig:correlation}.   
Due to the attractive nature of the two-body interaction   
used, the two-body density presents a clear correlation along the axis $(x_1+x_2)/2$ that is completely  
neglected at the Hartree-Fock level. Such a correlation is partially recovered when pairing   
is included in the   
HFB or BCS theory.   

\subsection{Some numerical aspects for dynamics with pairing}  
  
It is important to integrate the time-dependent equations of motion with  
a high-order method, because wave function conditions such as normalization  
and conserved quantities such as energy can be easy lost.    
The time scale for single-particle motion and direct reactions is  
several thousand of units of time, and we require numerical accuracy up to those  
times. For most of  
the results we present below, we have used the fourth-order Runge-Kutta  
algorithm (RK4).  The calculations in Ref. \cite{Ste11} on the other hand use  
a sixth-order Adam-Bashford algorithm, and we have tested that as well.  
  
Typically, we take a box of dimension $X_{max}=500$, giving the HFB matrices a dimension of  
$4X_{max}/\Delta x = 2000$. The single-particle Hamiltonian has a range  
up to $\sim 2$ , which requires a fairly small time step.  We  
take $\Delta t=0.263$.

\subsubsection{Ehrenfest vs Thouless equation of motion}  
  
As has been stressed in section  
\ref{sec:tdhfb}, the equation of motion on the $(u_\alpha, v_\alpha)$  
components are not unique.  We have implemented two of the formulations
below, namely the ``Ehrenfest" (Eq. (\ref{eq:wave1})) and the ``Thouless" 
( Eq. (\ref{eq:wave2})) equations of  
motion. 
The numerical integration can be carried out 
very accurately using each version of the equations.   
We found that the Thouless equation has a better precision  
than the Ehrenfest equation using RK4 at a fixed time step.  
However, it  
turns out that the Ehrenfest formulation is three or four times faster  
than the Thouless one, due to the smaller number of matrix operation in Eqs.  
(\ref{eq:wave1}) compared to Eqs. (\ref{eq:wave2}).  
Since the computational time is a crucial   
aspect of the numerical treatment, the  
standard Ehrenfest equation is a better choice. 
The density formulation 
(Eq. (\ref{eq:evol_rho}-\ref{eq:evol_kappa})) would have a similar number of matrix operations to the
Thouless formulation, but we have not investigated the numerical performance
of this third alternative.

\subsubsection{Imaginary absorbing potential}

Particle loss is monitored by computing the number of particles  
having $|x| < X_0/2$.  Particles can be reflected from the edges of  
the box and obscure this measure of evaporation, so we have to add  
an absorbing potential $h_i$ near the edges.  As mentioned in Ref. 
\cite{Ave08}, the specific form of the 
absorbing potential is not obvious, because it should decrease 
the particle number without affecting the normalization of the wave 
function.   
It can be shown that the following prescription satisfies these 
requirements, 
\begin{eqnarray}  
i \hbar \frac{\partial}{\partial t}   
\left(  
\begin{array} {c}  
 u \\  
 v   
\end{array}  
 \right)  
& = &   
\left(  
\begin{array} {cc}  
h - \rho h_i & -\Delta-\kappa h_i   \\  
-\Delta^*+\kappa^*h_i  & -h^*+(1-\rho^*)h_i   
\end{array}  
 \right)  
 \left(  
\begin{array} {c}  
 u \\  
 v   
\end{array}  
 \right).  
\label{eq:evol_uv2_im} \nonumber  
\end{eqnarray}  
In particular, the above equation preserves the unitarity property   $u u^\dagger + v^* v^t = 1$. \\  
In applications below, the imaginary potential is taken as  
\begin{eqnarray}  
h_i(x) &=& 0 ~~~ {\rm for} ~ |x| ~ \langle ~(X_{max}-x_{im}),\nonumber\\  
h_i(x) &=& iV_{im}\frac{|x|-X_{max}+x_{im}}{x_{im}}~~~ {\rm for} ~ |x| ~\rangle~ (X_{max}-x_{im}) \nonumber  
\end{eqnarray}    
with $X_{max}=L_{\rm max}/2$, $V_{im}=-7.716 \times 10^{-3}$ and $x_{im}=37.5$. \\  
We have compared TDHFB evolution in small box including the imaginary 
potential with the corresponding evolution
in very large box to check that the present method is a practical way to
suppress the reflected particles.
We also found that a simplier prescription is also adequate for our
purposes.  Namely, one can apply the imaginary potential to the $v$
amplitudes alone, with the equation of motion
\begin{eqnarray}  
i \hbar \frac{\partial}{\partial t}   
\left(  
\begin{array} {c}  
 u \\  
 v   
\end{array}  
 \right)  
& = &   
\left(  
\begin{array} {cc}  
h  & -\Delta  \\  
-\Delta^*  & -h^*+h_i   
\end{array}  
 \right)  
 \left(  
\begin{array} {c}  
 u \\  
 v   
\end{array}  
 \right)  
\label{eq:evol_uv2_im2}. \nonumber  
\end{eqnarray}     
This prescription violates unitarity, but the results using it
could not be distinguished from the correct evolution.

\subsection{Particle evaporation}  
  
\begin{figure}[htbp]  
\includegraphics[width=1\linewidth]{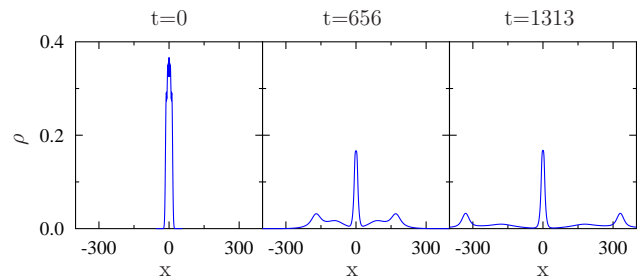}  
\caption{Evolution of the local one-body density $n(x)$ of a system of $N=10$ particles. The system   
is initially confined in a harmonic trap. At $t \ge 0$, the external constrained is relaxed.   
}  
\label{fig:evap}  
\end{figure}

To simulate an evaporating system, we start with a wave function that  
is constrained to be largely inside the potential well $U(x)$.  This is  
achieved by adding a small harmonic constraining field $\lambda r^2$ to   
the Hamiltonian and solving for the HFB ground state. At time $t \ge 0$, the harmonic   
constraint is removed inducing a monopole oscillation of the system that is eventually damped out by particle evaporation.  
This is illustrated in Fig. \ref{fig:evap}, showing snapshots of the   
density at different times with the system evolved with the TDHF equations  
of motion, ie. without pairing.  

 \subsubsection {Comparison between the exact solution and TDHFB} 

In this section we will compare the particle emission of TDHFB with
that given by the two-particle Schr\"odinger equation, solved numerically.
One should not expect close agreement under all conditions for two
reasons.  The total emission probability in the final state can be
calculated easily in the Schr\"odinger dynamics by taking the overlap
of the initial state with the bound solutions.  The TDHFB dynamics
on the other hand may have no binding when the average particle number
on the nucleus becomes small.

It is also not possible to set 
the initial conditions for the HFB wave function to correspond exactly
to the two-particle wave function of the Schr\"odinger equation; one
sees this already in Fig. 2.  As described above, the initial state for
the Schr\"odinger equation is squeezed ground state, namely the lowest 
state of the two-particle system in the presence of a harmonic external 
potential.  A  corresponding HFB wave function could be constructed
by using the BCS form of the wave function and
requiring that it have the same one-particle density matrix.  This
turns out to not work well, due to high momentum components in the
wave function that are not properly controlled by the HFB pairing
field.  We found that a better prescription is to make a corresponding
squeezed ground state in the HFB treatment.  We use this prescription
for the comparison shown below.

\begin{figure}[htbp]  
\includegraphics[width=1\linewidth]{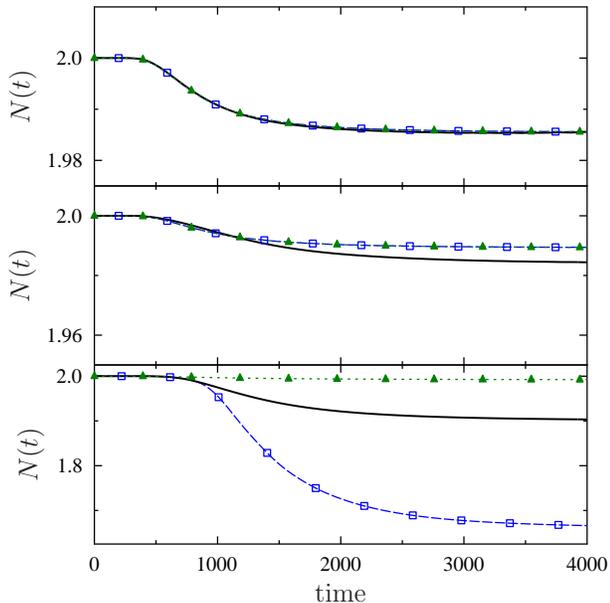}   
  \caption{Number of particles evaporated from an initially compressed system
  with initially $N=2$ as a function of time obtained with the exact (solid
black line) and TDHF (filled green triangle) and TDHFB (open blue squares).  Results with different two-body
  interaction strengths (case (a), (b) and (c)) are respectively shown from
  top to bottom (see text).
}  
  \label{fig:N2}  
 \end{figure}

We measure the number of particles inside the system by the
quantity 
 \begin{eqnarray}  
N(t) & = & \int_{|x|\langle X_{\rm box}} 2 n(x,t)dx,  
\end{eqnarray}   
where $X_{\rm box}$ here is taken as $100$. Note that the system is centered
at $x=0$.
The evolution of $N(t)$ for several cases is shown in
Figure \ref{fig:N2}, comparing the exact results with the HFB approximation.

In the case (a) and (b), TDHF and TDHFB are identical. Indeed, the minimization 
of HFB equation to get the initial state leads to a pure Slater determinant state. 
In these case, the TDHF evolution is very close to the exact solution. Note that, in
this regime, the evaporation is dominated by the mean-field contribution and
pairing has a weak effect on particle emission.  As the interaction strength
increases, the TDHFB and TDHF results starts to deviate from each other as well as
from the exact evolution. \textcolor{black}{As the interaction strength increases, the role of pairing 
and, more generally, correlations
on evaporation becomes more important. The TDHF evolution largely underestimate 
the emission in case (c). This stems from the fact that mean-field is not able to properly describe
the diffusion of the occupation probability around the Fermi energy in the initial state and the dynamical 
scattering of single-particles during the evolution induced by correlations beyond the Hartree-Fock.
In bottom panel of figure \ref{fig:N2}, the lack of evaporation in TDHF is due to the fact that all initial 
occupied states can be decomposed onto bound states of the corresponding mean-field. A similar situation 
occurs for the $N=10$ case presented below. 
}
 
A precise study of the strongest coupling case (case (c)),  
which is the only case above the HFB threshold for the initial state, 
shows that the time scale associated to particle evaporation is properly accounted for in TDHFB. This could indeed  
be seen in bottom part of figure \ref{fig:N2} where we see that the time at which $N(t)$ starts to decrease is the same in 
the exact and in the TDHFB case. This shows that the time-scale associated with the evaporation 
process is the same in the exact and TDHFB case. In the long time limit, TDHFB overestimates the average number of emitted 
particles.  
Accordingly, it could be anticipated that the internal motion of the system is more damped in the latter case than  
in reality. We indeed have checked that the damping width of the monopole resonance is larger in TDHFB compared  
to the exact solution.   
 
It should be noted that the approximation leading to TDHFB can only be justified for the short-time evolution.
Indeed, even starting from a quasi-particle
state, correlation beyond TDHFB might built up in time, like for instance
four quasi-particle excitations.
    
\subsubsection {Comparison between the exact solution and TDHF+BCS} 
  
\begin{figure}[htbp]  
\includegraphics[width=1\linewidth]{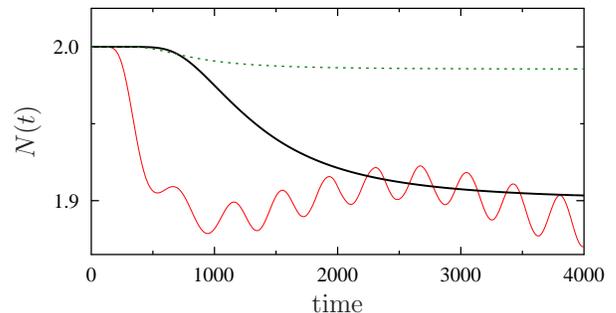}   
  \caption{Number of particles evaporated from an initially compressed system with initially $N=2$ as a function of time  
  obtained with the exact (solid black line), TDHF (dashed green line) and TDHF+BCS theory (thin red line) in the case of parameter set (c).  }  
  \label{fig:N2BCS}  
 \end{figure}

Here, the results obtained by using the TDHF+BCS equation of motion discussed in section \ref{sec:tdhfbcs}  
are presented. In figure \ref{fig:N2BCS}, an illustration of the result obtained in the  
case (c) is shown.   
   
\begin{figure}[htbp]  
\includegraphics[width=8.cm]{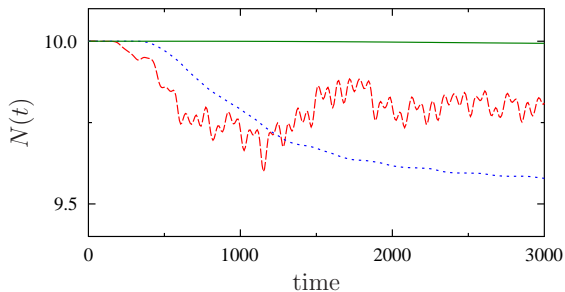}  
  \caption{Number of particles evaporated from an initially compressed system with initially $N=10$ as a function of time  
  obtained with the TDHF (solid line), TDHFB (dotted line) and TDHF+BCS 
theory (dashed line). 
 }  
  \label{fig:N10}  
 \end{figure}

Independently of the set of parameters used in the model case, it is generally observed that the TDHF+BCS   
theory leads to unrealistically fast early emission of particles compared to the exact case.   
This fast emission seems to be a generic feature   
of the BCS approach as illustrated in Figs. \ref{fig:N10} where $N=10$ particles are considered.   

Independently of the set of parameters used in the model case, it is
generally observed that the TDHF+BCS theory leads to unrealistically fast
early emission of particles compared to the exact case. This fast emission
seems to be a generic feature of the BCS approach as illustrated in Figs. \ref{fig:N10} 
where N = 10 particles are considered. This observed fast emission is very
likely connected the problem of applying BCS when continuum states are
present in the wave functions .  The BCS ground state
has a unphysical gas of particles in the continuum
rather than an exponential decay into the vaccum \cite{Dob84}.
This was one of the
historical reason why HFB was preferred to BCS in nuclear structure studies.
It is of course possible to reduce the continuum problem by truncating
the number of single-particle states that contribute to
pairing.  However, we do not know any systematic way to carry this out
without reference to more reliable calculational methods. 

In studies dedicated to nuclear structure, this is generally   
circumvented by reducing significantly the number of single-particle states that contribute to pairing. Then, only states   
with single-particle  energy within a given range $\Delta E$ around the Fermi energy are used, where $\Delta E$ is of the order of few MeV.  In Figures \ref{fig:N2}-\ref{fig:N10}, this restriction has not been made and a large set of single-particles is retained.  \textcolor{black}{If the energy window $\Delta E$ is reduced, the time-scale associated to particle evaporation is increased and eventually becomes more consistent with the exact dynamics. Conjointly, the asymptotic number of evaporated particles is significantly reduced and approaches the TDHF case 
as $\Delta E$ goes to zero.} 
  
 It should be mentioned that in realistic three-dimensional calculations, there is no flexibility   
 in the selection of single-particle states contributing to the dynamics. Indeed, static calculation  
 are already made with a specific choice of single-particle space in such a way that with an effective   
 force in the pairing channel, the gap has a reasonable value. Accordingly, the dynamics should be   
 made with the same set of single-particles states has is already done in Ref. \cite{Eba10}. 
   
   
\subsubsection{Spurious oscillation in TDHF+BCS theory}  
  
 \label{sec:spurious}

In the long time evolution, oscillation of the number of particles, similar to those displayed in figure   
\ref{fig:intro1} are observed in TDHF+BCS, see Figs. \ref{fig:N2BCS} and \ref{fig:N10}.  
Such oscillations are   
absent in the TDHFB theory.  From the application presented here, we can conclude that the spurious oscillations  
is a generic effect in TDHF+BCS. It occurs even if a finite range interaction is used. Finally, this   
problem is solved when TDHFB is used.   
  
\begin{figure}[htbp]  
\includegraphics[width=9.cm]{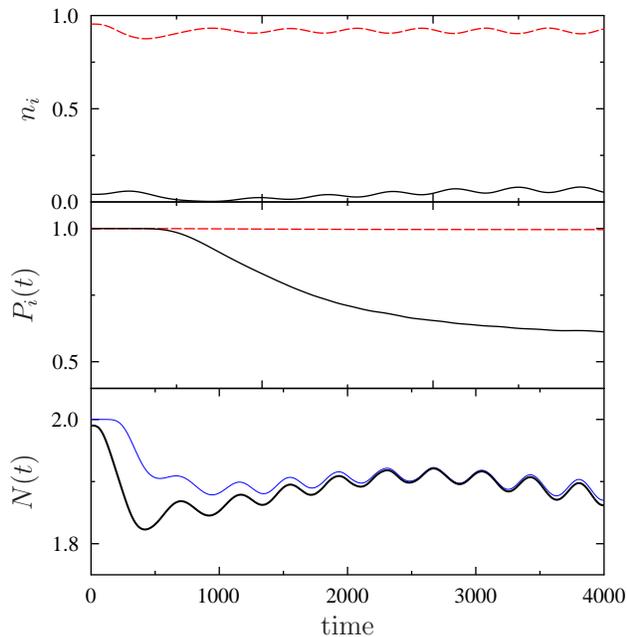}  
  \caption{ Top: Evolution of occupation numbers  of the two closest states above (state 2, dashed line) and below (state 1, solid line) the Fermi energy as   
  a function of time obtained in TDHF+BCS (parameters set (c)).   
  Middle: Evolution of the corresponding portion of the wave-function remaining inside the box.   
  Bottom: Evolution of $N(t)$ (thin line) and of $N'(t)$ (thick line)  as a function of time.}  
  \label{fig:nit}  
 \end{figure}

To better characterize the oscillation, $N(t)$ can be expressed in the canonical basis as
 \begin{eqnarray}  
N(t) = \sum_i n_i(t) P_i(t)  ,
\end{eqnarray}  
where $n_i(t)$ and $P_i(t)$ denote respectively the occupation numbers and the probability of the canonical orbital $i$ 
inside the box:  
\begin{eqnarray}  
P_i(t) = \int\limits_{|x| \langle X_{\rm box}} |\varphi_i(x,t)|^2 dx  .
\end{eqnarray}  
An illustration of $N(t)$ for the two particle case (c) is given in figure \ref{fig:nit}.  
The observed evolution is mainly due to the evolution of the two closest levels below and   
above the particle emission threshold labelled respectively by "$1$" and "$2$".  These two levels   
verify  $n_1(t) + n_2(t) \simeq 1$.
During time evolution, the unbound level is continuously emitted while the bound level remains in the box, i.e.   
$P_1(t) = 1$. Assuming, that only these two levels contribute to the particle emission, an estimate $N'(t)$ of the 
number of evaporated particles is given by \footnote{Note that the factor 2 here comes from the initial degeneracy}:    
\begin{eqnarray}  
N'(t) & = & 2 \left[n_1(t) P_1(t) + n_2(t) P_2(t) \right]  .
\end{eqnarray}  
The evolution of $n_i(t)$ and $P_i(t)$ for $i=1,2$, as well as $N'(t)$ are shown in Fig. \ref{fig:nit} attesting  
for the validity of the two-level approximation. As seen in bottom part of this figure, $N'(t)$ is very close   
from its exact value $N(t)$ and oscillations are due to oscillations in occupation numbers

\begin{figure}[htbp]  
\includegraphics[width=9.cm]{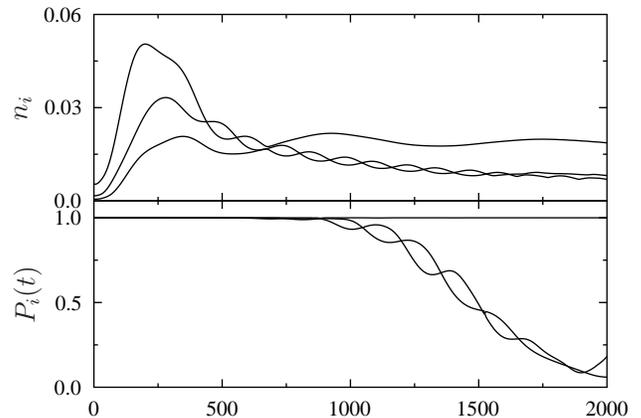}  
  \caption{ Top: Evolution of occupation numbers of the three main single-particle canonical states contributing   
  to the particle evaporation for the exact dynamics. The corresponding values of $P_i(t)$ are shown in the bottom part. }  
  \label{fig:nitexact}  
 \end{figure}

Such oscillations of  occupation numbers are expected in any theory beyond TDHF, including the TDHFB and/or exact evolution (see Fig. \ref{fig:nitexact}). However, these   
 theories do not lead to unphysical evolution of particle number.  
 The difference between TDHF+BCS and the two other theories   
 stems from the approximation made to get the equation of motions. Indeed, by neglecting the off-diagonal matrix elements   
 of the pairing field, the single-particle evolution reduces to self-consistent mean-field dynamics, similar to the TDHF one.
 The effect of correlation only enters into the occupation numbers evolution, and 
only affects the single-particle evolution through the density dependence of the self-consistent mean-field.   
   
Usually, correlation is expected to induce a mixing of single-particle states. Indeed, the evolution of the one-body   
 density matrix in the presence of correlation is given by:  
\begin{eqnarray}  
i\hbar \frac{\partial \rho}{ \partial t} & = & [h(\rho), \rho] + Tr_2 [v_{12}, C_{12}]   , 
\end{eqnarray}  
where $h(\rho)$ is the mean-field of the correlated state while $v_{12}$ and $C_{12}$ denotes the two-body interaction   
and correlation matrix respectively (see Ref. \cite{Lac04} for more details).  In both TDHFB and exact solution, the second term   
induces an extra mixing of single-particle states that is neglected in TDHF+BCS. It turns out that this mixing is essential to   
compensate the possible oscillations in occupation numbers. This is clearly illustrated in Fig. \ref{fig:nitexact}  where 
the quantity $P_i(t)$  are shown to oscillate coherently with $n_i(t)$ in the exact case (similar behavior is observed in TDHFB
evolution).   
  
\subsubsection{Link with the break-down of continuity equation in TDHF+BCS}  
Starting from the expression (\ref{eq:contbcs}) derived for TDHF+BCS, the evolution of particle   
number inside the box is given by  
\begin{eqnarray}  
\frac{d N(t)}{dt}  & = & - \int\limits_{|x| \langle X_{\rm box}} {\rm div} (j (x,t))dx \nonumber \\  
&+& \sum_i  P_i(t) \left( \frac{dn_i (t)}{dt} \right)  .
\label{eq:contbcs2}  
\end{eqnarray}     
Introducing two sets of real functions $R_i(x,t)$ and $S_i(x,t)$ for each wave-packet such that:  
\begin{eqnarray}  
\varphi_i(x,t) = R_i(x,t) \exp(i S_i(x,t)/\hbar)  ,
\end{eqnarray}  
and making use of partial integration technique, the first term in eq. (\ref{eq:contbcs2}) can be recast as:  
\begin{eqnarray}  
\int\limits_{|x| \langle X_{\rm box}} {\rm div} (j (x,t))dx &=& 2\sum_i n_i  |\varphi_i(X_{\rm box},t)|^2 v_i(X_{\rm box},t), \nonumber   
\end{eqnarray}  
where $v_i$ denotes the local velocity of the particle defined through $v_i(x,t) \equiv  \nabla S_i (x,t)/m$.   
  
This term is the expected physical term expected to appear in any well defined transport theory  
that relates the number of particles inside the box to the flow of particles outgoing at the   
boundary of the box. However, due to the presence of the second term in eq. (\ref{eq:contbcs2}), oscillation  
of occupation numbers that are not compensated by oscillation of the probability $P_i(t)$ (see figure \ref{fig:nit})  
lead to spurious behavior of the particle number.  The only way out to avoid this problem in a TDHF+BCS approach   
is to freeze the occupation numbers during the evolution.  
    
 \subsubsection{TDHF+BCS with frozen occupation numbers}  
    
 To incorporate pairing in a transport model we are facing the difficulty that the   
 TDHFB theory is very demanding numerically. A possible solution to this difficulty, would be   
 to use the simpler TDHF+BCS approach. However, in view of preceding sections, the approximation  
 made to obtain TDHF+BCS leads to unphysical behavior especially when continuum plays a significant   
 role: strange behavior of particle emission, gas problem.   
   
 We have seen in section \ref{sec:spurious}, that the pathologies of TDHF+BCS   
 comes from the evolution of occupation that should normally be accompanied by a consistent   
 mixing of the single-particle states along the dynamical path.  This approximation does not seem  
 to be critical in the study of static properties of nuclei and most often, for not too exotic nuclei, BCS theory   
 provides a fairly good approximation to HFB.      
   
\begin{figure}[htbp]  
  \includegraphics[width=9.cm]{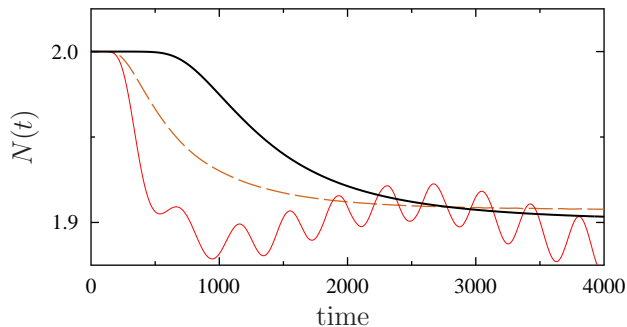}  
  \caption{ Evolution of the number of particles evaporated from an initially compressed system of $N=2$ particles. The exact result  
  (thick solid line) is compared to the TDHF+BCS with (dashed line) and without (thin solid line) the frozen occupation number   
  approximation. The simulation has been made with the same parameters set as the lower panel of figure \ref{fig:N2}.}  
  \label{fig:p2_ni_cst_bcs}  
 \end{figure}

 A simple prescription to avoid non-physical evolution in TDHF+BCS is to assume that the occupation numbers are frozen   
 during the time-evolution, this approximation is called hereafter {\it frozen occupation approximation} (FOA).   An illustration   
 of the FOA effect on particle evaporation is shown in figure \ref{fig:p2_ni_cst_bcs} (dashed line).    
As anticipated, spurious oscillations of the particle number evaporation disappear in the FOA.  It turns out, that, for   
the specific set of parameters used in the example of figure \ref{fig:N2}, the asymptotic number of particle evaporated   
is in very good agreement with the exact case, much better than the TDHFB solution (see figure \ref{fig:N2}). However,   
the agreement depends on the parameters that are used and not systematic conclusion can be drawn.  
Note that this approximation has already been used in realistic calculation for example to study dipole giant resonances \cite{Mar05}. 
   \section{Summary}  
  
In this article, different transport theories able to incorporate pairing are discussed. One important conclusion   
is that theories like TDHF+BCS where the continuity equation is not respected can lead to unphysical results.  
More specifically, the effect of pairing on particle emission has been analyzed here using a simple one-dimensional   
Hamiltonian that can be solved exactly for the case of two particles.   
From the systematic study we have made by changing the interaction strength and/or particle number, pairing does affect   
significantly the particle number emission.    
While the TDHF approach generally underestimate significantly the number of emitted   
particles, an enhancement of particle evaporation is observed when pairing is included.   
This effect is automatically   
included in both TDHFB or TDHF+BCS theory. Only TDHFB provides a good description 
of particle emission at short time but might deviates from the exact dynamics 
at longer time due to accumulated correlation effects beyond this approach.

While the asymptotic number of emitted particles is quite reasonable, 
TDHF+BCS leads to unphysical rapid emission and spurious oscillations of the number 
of emitted particles. The direct use of TDHF+BCS, that would be highly desirable from   
the practical point of view, is plagued with unphysical behavior, and, as we have shown, it is preferable to use a simplified version   
where occupation numbers are frozen to their initial value.   
    
In summary, both TDHFB and TDHF+BCS with constant occupation numbers can eventually   
be used to describe a physical system while TDHF+BCS with varying occupation numbers should be avoided.   
While TDHFB is expected to have a reacher dynamics, due to its simplicity,   
the second transport theory remains quite attractive. 
  
\section*{Acknowledgment}  
  GFB thanks A. Bulgac for discussions.  We also acknowledgment the  
France-US Institute for Physics of Exotic Nucleus for collaborative  
support.  GFB was also supported by the US Dept. of Energy under  
Grant DE-FG02-00ER41132.

\end{document}